\documentclass[
aip,
amsmath,amssymb,
reprint,%
]{revtex4-1}

\pdfoutput=1
\usepackage{isomath}
\usepackage{amsmath,amsthm}
\usepackage{amsbsy}
\usepackage{amssymb}
\usepackage{amscd}
\usepackage{amsfonts}
\usepackage{stmaryrd}
\usepackage{siunitx}
\usepackage{euscript}
\usepackage[utf8]{inputenc}
\usepackage[T1]{fontenc}
\usepackage{newtxtext} 
\everymath{\displaystyle}
\usepackage{exscale}

\usepackage{graphicx}
\usepackage{boxedminipage}
\usepackage{calc}
\usepackage[usenames,dvipsnames]{xcolor}
\graphicspath{ {media/} }

\usepackage{setspace}
\usepackage{enumitem}
\setitemize{noitemsep,topsep=0pt,parsep=0pt,partopsep=0pt}
\setenumerate{noitemsep,topsep=0pt,parsep=0pt,partopsep=0pt}
\setdescription{noitemsep,topsep=0pt,parsep=0pt,partopsep=0pt}

\usepackage[colorinlistoftodos, color=green!40,prependcaption]{todonotes}

\usepackage{soul} 

\usepackage{xr-hyper}
\makeatletter
\newcommand*{\addFileDependency}[1]{
  \typeout{(#1)}
  \@addtofilelist{#1}
  \IfFileExists{#1}{}{\typeout{No file #1.}}
}
\makeatother

\usepackage[,colorlinks,urlcolor=blue,citecolor=black]{hyperref}
\usepackage[normalem]{ulem}

\setuptodonotes{inline}
\usepackage{isomath}
\usepackage{amsmath}
\usepackage{amssymb}
\usepackage{amscd}
\usepackage{amsfonts}

\newcommand{\parderiv}[2]{\frac{\partial #1}{\partial #2}}
\newcommand{\dm}{\ \mathrm{d}}
\newcommand{\deriv}[2]{\frac{\dm #1}{\dm #2}}

\usepackage{dcolumn}
\usepackage{caption}
\usepackage{subcaption} 

\usepackage{csquotes}
\usepackage{svg}
\usepackage[symbol]{footmisc}
\usepackage{scalerel}
\usepackage{tikz}
\usetikzlibrary{svg.path}

\definecolor{orcidlogocol}{HTML}{A6CE39}
\tikzset{
	orcidlogo/.pic={
		\fill[orcidlogocol] svg{M256,128c0,70.7-57.3,128-128,128C57.3,256,0,198.7,0,128C0,57.3,57.3,0,128,0C198.7,0,256,57.3,256,128z};
		\fill[white] svg{M86.3,186.2H70.9V79.1h15.4v48.4V186.2z}
		svg{M108.9,79.1h41.6c39.6,0,57,28.3,57,53.6c0,27.5-21.5,53.6-56.8,53.6h-41.8V79.1z M124.3,172.4h24.5c34.9,0,42.9-26.5,42.9-39.7c0-21.5-13.7-39.7-43.7-39.7h-23.7V172.4z}
		svg{M88.7,56.8c0,5.5-4.5,10.1-10.1,10.1c-5.6,0-10.1-4.6-10.1-10.1c0-5.6,4.5-10.1,10.1-10.1C84.2,46.7,88.7,51.3,88.7,56.8z};
	}
}

\newcommand\orcidicon[1]{\href{https://orcid.org/#1}{\mbox{\scalerel*{
				\begin{tikzpicture}[yscale=-1,transform shape]
					\pic{orcidlogo};
				\end{tikzpicture}
			}{|}}}}

\usepackage{verbatim}

\newcommand{\etal}{\textit{et al.}}
\newcommand{\eff}{\varepsilon_\mathrm{eff}}
\newcommand{\Qnew}{Q_\mathrm{new}}
\renewcommand{\Re}{\operatorname{Re}}
\renewcommand{\Im}{\operatorname{Im}}
\makeatletter
\def\@email#1#2{%
	\endgroup
	\patchcmd{\titleblock@produce}
	{\frontmatter@RRAPformat}
	{\frontmatter@RRAPformat{\produce@RRAP{*#1\href{mailto:#2}{#2}}}\frontmatter@RRAPformat}
	{}{}
}%
\makeatother

\begin{document}
	
	\preprint{AIP/123-QED}
	
	\title[Q-Factor Bounds]{Bounds on the Quality-factor of Two-phase Quasi-static Metamaterial Resonators and Optimal Microstructure Designs} 
	\affiliation{Department of Mathematics, University of Utah, Salt Lake City, UT 84112, U.S.A}
	\author{Kshiteej J. Deshmukh\orcidicon
		{0000-0002-6825-4280}\thanks{*}}
	
	\author{Graeme W. Milton\orcidicon{0000-0002-4000-3375}}%
	\email[Corresponding author:\ ]{kjdeshmu@math.utah.edu}
	
	
	
	\date{\today}
	\begin{abstract}
		Material resonances are fundamentally important in the field of nano-photonics and optics.
		So it is of great interest to know what are the limits to which they can be tuned. 
		The bandwidth of the resonances in materials is an important feature which is commonly characterized by using the quality (Q) factor.
		We present bounds on the quality factor of two-phase quasi-static metamaterial resonators evaluated at a given resonant frequency by introducing an alternative definition for the Q-factor in terms of the complex effective permittivity of the composite material.
		Optimal metamaterial microstrcuture designs achieving points on these bounds are presented.
		The most interesting optimal microstructure, is a limiting case of doubly coated ellipsoids that attains points on the lower bound.
		We also obtain bounds on Q for three dimensional, isotropic, and fixed volume fraction two-phase quasi-static metamaterials. 
		Some almost optimal isotropic microstructure geometries are identified.
	\end{abstract}
	\maketitle
	
	The explosion of interest in metamaterials has been driven by the realization that they can break expected limits.
	These limits, or bounds, are based on assumptions that do not hold for the metamaterials in question. 
	Naturally, one wants to know what these bounds are, and here we focus on a fundamental problem: bounding the Q-factor of resonances in metamaterials, under the assumption that one is in the quasi-static limit.

	Resonances in materials have led to many exciting properties and applications in nano-photonics and optics.
	A famous example of material resonance is the "Lycurgus cup", which is a $4^{th}$ century Roman drinking cup made of glass  with fine particles of gold suspended in it. 
	The resonances of the gold particles at optical wavelengths cause it to appear either red or green depending on where the light shines from.
	By making the gold particles hollow one can shift the resonant frequency\cite{aden1951scattering,averitt1997plasmon}, even into the infrared where nanoshells have proved significant in destroying cancer cells\cite{rastinehad2019gold}. 
	Resonances are also studied extensively in antenna theory. 
	A large body of work is available on the design of antennas \cite{gustafsson2012optimal, gustafsson2019tradeoff} with applications ranging from electronic sensors to radio-frequency energy harvesting, and wearable antennas \cite{qian1998progress, guo2016simultaneous,zhu2021strain}.
	An important factor in  antenna design is the operating frequency-bandwidth about the resonance, and this is typically quantified by defining the quality (Q) factor of an antenna \cite{collin1964evaluation}. 
	Depending on the application it may be desirable to have a low Q or a high Q.

	A number of definitions for Q are found in literature that try to best approximate the exact bandwidth of an antenna\cite{collin1964evaluation,fante1969quality,rhodes1976observable,yaghjian2005impedance,gustafsson2006bandwidth}.
	The two most common conventional definitions of Q-factor that have been in use are: one, the ratio of energy stored to energy radiated or dissipated; and, two, the ratio of center (resonance) frequency to frequency-bandwidth.
	\begin{figure*}[ht!]
		\centering
		\includegraphics[width=0.8\textwidth]{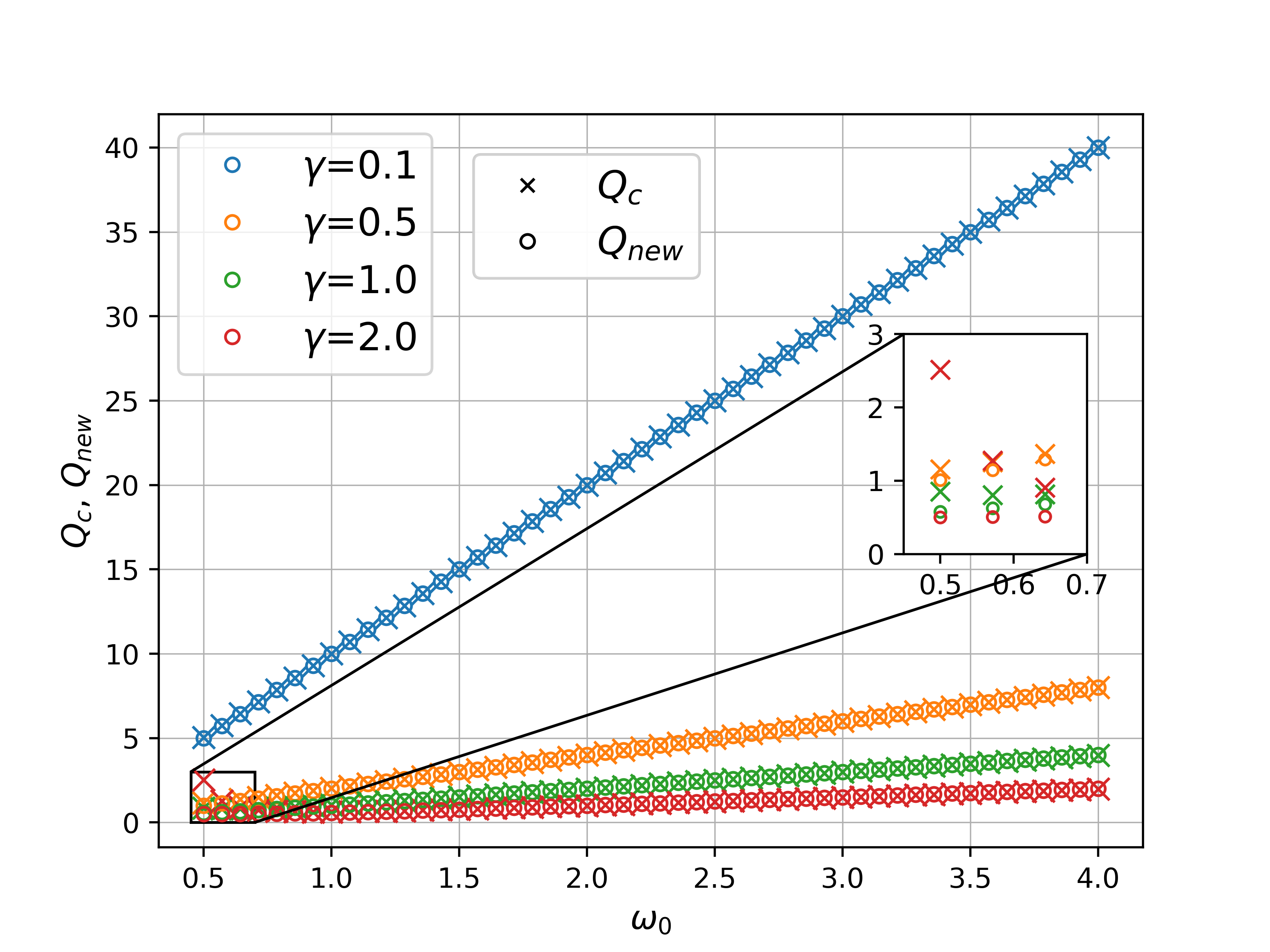}
		\caption{$Q_c$ (circle marker) vs. $\Qnew$ (cross marker) comparison for Lorentz models with varying values of the damping coefficient $\gamma$ and resonant frequency $\omega_0$. 
			In the limits $\omega_0/\gamma>>1$ and $\omega_0/\Delta\omega>>1$, we see good agreement between $Q_c$ and $\Qnew$. 
			The inset shows the discrepancy between $Q_c$ and $\Qnew$ near the origin where the approximation no longer holds.
		}
		\label{fig:Qc vs Qnew}
	\end{figure*}
	
	Using the first definition of Q-factor, there are several works that have proposed limitations or bounds on the Q-factor of antennas. 
	Chu\cite{chu1948physical} first set a lower limit on Q-factor for small antennas.
	Stuart\cite{stuart2008bandwidth} proposed bandwidth limitations on small antennas with negative permittivity materials, while  lower bounds on small antennas were given by Yaghjian \etal\cite{yaghjian2010lower}.
	Gustafsson \etal\cite{gustafsson2012optimal} considered optimizing current densities to numerically obtain physical bounds on antennas of arbitrary shape and size, and also provided shape dependent bounds in a series of papers\cite{gustafsson2012physical,gustafsson2015physical,jonsson2015stored}.
	More recently, lower bounds on Q-factors of small-size, high-permittivity, dielectric resonators were obtained by Pascale \etal\cite{pascale2023lower}.

	In this work, we focus on the problem of finding bounds on the Q-factor in lossy two-phase quasi-static metamaterial resonators, and on identifying microgeometries that achieve those bounds. 
	The use of metamaterials enriches the design space of resonators.
	To our knowledge, bounds on Q in this case have not been studied so far. 
	We consider two-phase metamaterials where the 2 pure phases are isotropic, dielectric materials, with complex relative permittivity $\epsilon_1(\omega)$ of one phase given, for example, by a Drude model or a Lorentz model, and the relative permittivity $\epsilon_2(\omega)=\epsilon_2$ of the other phase is taken to be a real constant. 
	The effective permittivity of the metamaterial is some complex valued function of the permittivities of the pure phases $\varepsilon_\mathrm{eff}(\omega) = \varepsilon_\mathrm{eff}(\epsilon_1(\omega), \epsilon_2)$. 
	Without loss of generality,  we can rescale the dimensions and set the permittivity of the constant  phase as\cite{milton1981bounds}, $\epsilon_2=1$.
	Considering the ambiguity associated with defining the energy stored in a lossy material\cite{gustafsson2015stored, schab2018energy}, we adopt the second definition for Q, i.e., $Q=\omega_R/ \Delta\omega$, where $\omega_R$ is the resonance frequency and $\Delta \omega$ is the bandwidth at half-height of the resonance value, and refer to this definition as the  conventional Q-factor, $Q_c$. 
	However, to proceed with finding the bounds on Q in two-phase metamaterials it is desirable to have an expression for Q in terms of the material parameters at the resonant frequency. 
	
	Assume that the response of a material permittivity is given by a Lorentz model,
	\begin{equation}\label{Lorentz model}
		\varepsilon = 1 +  \frac{\omega_p^2}{\omega_0^2 - \omega^2 -\imath\omega \gamma}, 
	\end{equation}
	where $\omega_p$, $\omega_0$, and $\gamma$ are the plasma frequency, natural frequency, and damping coefficient respectively.
	We show analytically, that in the limit $\omega_0/\gamma>>1$ (see supplementary information (SI) Section \textcolor{red}{S1} and Figure \textcolor{red}{S1}   for details), $\gamma\approx\Delta\omega$, i.e. the bandwidth is narrow compared to the resonance frequency, and the resonant frequency $\omega_R$ approaches the natural frequency $\omega_0$ ($\omega_R\approx\omega_0$). 
	Then, the expression for $Q_c$ at resonance can be approximated alternatively by defining a new Q-factor ($\Qnew$) in terms of the material parameters:
	\begin{equation}\label{Q_new}
		Q_c \approx \Qnew = 
		-\frac{\omega_0}{2}
		\left[ 
		\Im\Big(\varepsilon(\omega_0)\Big)\right]^{-1}
		\Re\Big(\deriv{\varepsilon(\omega)}{\omega}\Big)\Big\vert_{\omega=\omega_0}.
	\end{equation}
	Here $\Re(\cdot)$ and $\Im(\cdot)$ denote the real and imaginary parts.
	This expression \eqref{Q_new} is similar to  ones found in the literature for antennas with known impedance\cite{rhodes1976observable,yaghjian2005impedance}.
	Figure \ref{fig:Qc vs Qnew} illustrates the good agreement between $Q_c$ and $\Qnew$ for different Lorentz models with varying values of $\gamma$ and $\omega_0$ even for moderately large values of $\omega_0/\gamma$. 
	The zoomed inset shows that close to origin, where $\omega_0$ is comparable to $\gamma$ and we are well outside the validity of the approximation, $Q_c$ and $\Qnew$ differ significantly.

	The main idea to obtain bounds on $\Qnew$ in two-phase metamaterials at a given resonance frequency ($\omega_R$) of the composite, is to correlate, at $\omega=\omega_R$, the quantities $\Re\Big(\deriv{\eff(\omega)}{\omega}\Big)$ and $\Im\Big(\eff(\omega)\Big)$ when  $\Im\Big(\deriv{\eff(\omega)}{\omega}\Big)=0$, the latter being a necessary condition for resonance at $\omega_R$.
	Note that the effective permittivity of the metamaterial is a function of its pure phases, $\epsilon_1(\omega)$ and $\epsilon_2=1$. 
	The derivative with respect to frequency of the effective permittivity can be rewritten as, 
	\begin{equation}
		\deriv{\eff(\omega)}{\omega}\Big\vert_{\omega=\omega_R}= \left.\bigg(\parderiv{\eff(\epsilon_1(\omega), 1)}{\epsilon_1} \deriv{\epsilon_1(\omega)}{\omega}\bigg)\right\vert_{\omega=\omega_R},
	\end{equation}
	where $\deriv{\epsilon_1(\omega)}{\omega}$ is a known quantity as $\epsilon_1(\omega)$ is known.
	\begin{figure*}[ht!]
		\centering
		\includegraphics[width=0.8\textwidth]{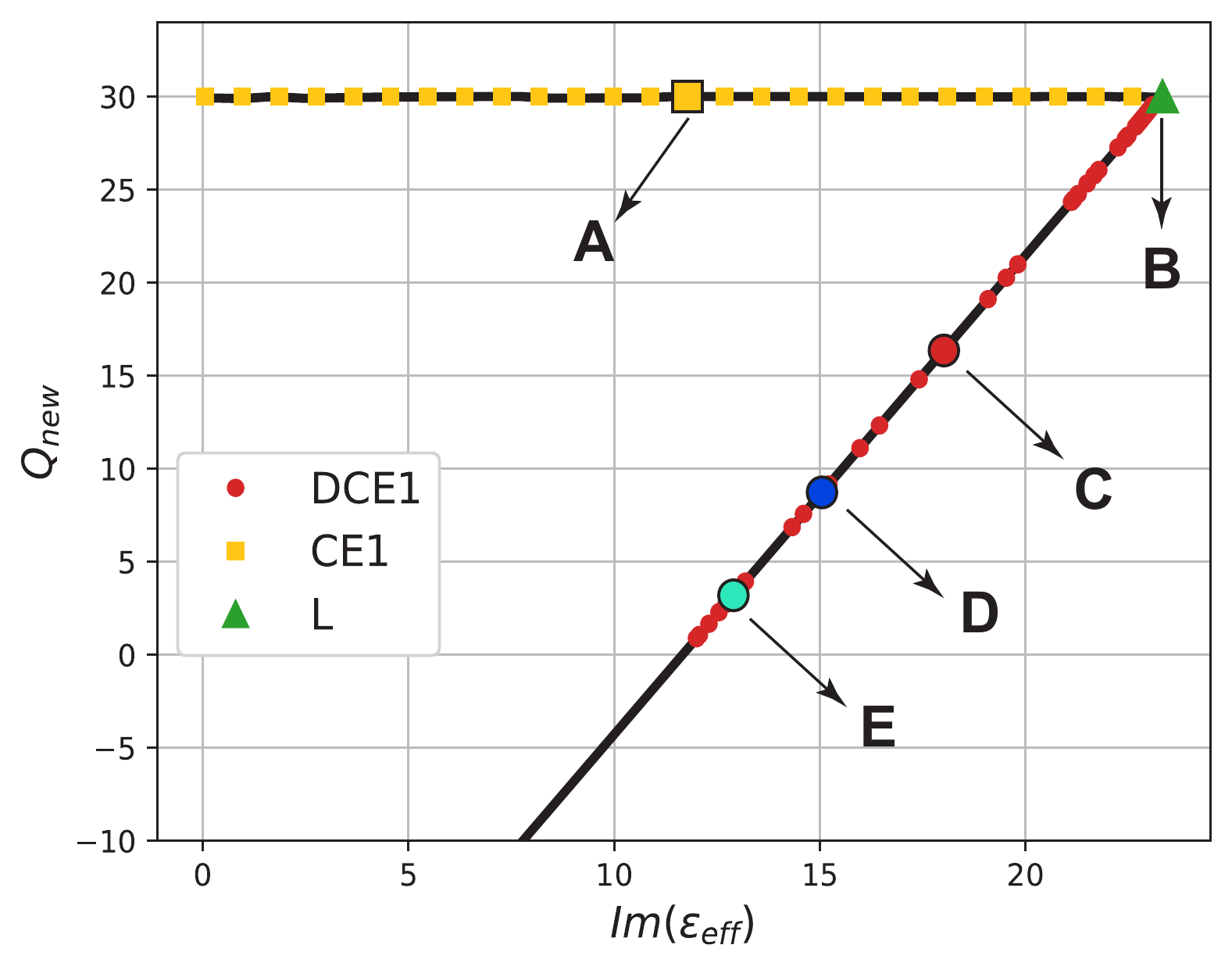}
		\caption{$\Qnew$ vs. $\Im(\eff)$  for two-phase metamaterial quasi-static resonators with $\epsilon_1(\omega)$ given by the Drude model.
			Bounds on $\Qnew$ (solid black curve) are shown as  the value of $\Im(\eff(\omega_R))$ at resonance frequency $\omega_R = 3$ is varied within the range of $\eff(\omega_R)$. 
			Metamaterial geometries achieving all points on these bounds are shown:
			Coated ellipsoids (CE1: yellow squares) attain values on the upper bound for $\Qnew$ as the eccentricities and volume fractions of the core phase are varied;
			laminate geometry (L: green triangle) attains the point on extreme right;
			points on the lower bound are achieved by doubly coated ellipsoids (DCE1: red circles).
			Specific points on the bounds are highlighted as A, B, C, D, and E and more information on the microstructure geometries at these points is given in Figure \ref{fig:Optimal designs}
		}
		\label{fig:Drude bounds}
	\end{figure*}
	This reduces the problem to correlating the quantities $\partial\eff(\epsilon_1(\omega), 1)/\partial\epsilon_1$ and $\Im\Big(\eff(\epsilon_1(\omega),1)\Big)$ at $\omega=\omega_R$. 
	To obtain this correlation we formulate the following problem: Given a fixed realizable value of $\Im\left(\eff(\epsilon_1(\omega),1)\right)$ at $\omega=\omega_R$, what are the bounds on the values of $\Re\Big(\partial\eff(\epsilon_1(\omega), 1)/\partial\epsilon_1\Big)$, with the constraint that $\left.\Im\bigg(\parderiv{\eff(\epsilon_1(\omega),1)}{\omega}\bigg)\right\vert_{\omega=\omega_R}=0$, which is associated with resonance at $\omega=\omega_R$? 
	We solve this problem numerically by using the bounds of Milton \etal\cite{milton1980bounds,milton1981bounds,milton1981bounds-b, milton2002theory} (see, Chapter 27 in Milton\cite{milton2002theory}), and more details on the method can be found in SI Section \textcolor{red}{S2}.
	In fact, in a broader mathematical context outside the theory of composites, there is a long history of such bounds: see, for example, Krein and Nudelman\cite{kreinnudelman}.
	From the bounds, we then easily obtain constraints on $\Qnew$ for any possible value of $\Im\Big(\eff(\epsilon_1(\omega_R),1)\Big)$.

	First, we present our results for 2-phase composites with $\epsilon_1(\omega)$ given by the Drude model,
	\begin{equation}\label{Drude model}
		\epsilon_1(\omega) = 1 + \omega_p^2/(\omega^2+\imath\omega\gamma).
	\end{equation}
	with, $\omega_p = 5$, $\gamma=0.1 $. 
	Plots for bounds on the effective complex permittivity $\eff(\omega_R)$ and for the corresponding range of   $\dm\eff/\dm\omega$ in this case are shown in the SI (Section \textcolor{red}{S2}). 
	Figure \ref{fig:Drude bounds} shows the wedge-shaped bounds on $\Qnew$ 
	(shown by the solid black curve) as the resonance value of $\Im(\eff)$ varies within the range prescribed by $\eff(\omega_R)$.
	To obtain our bounds on $\Qnew$, we allow for values of $\Re\Big(\dm{\eff(\omega)}/\dm{\omega}\Big)$ at $\omega=\omega_R$ that are positive. 
	Due to this and from \eqref{Q_new}, we see that  negative values of $\Qnew$ are allowed by the bounds as seen in Figure \ref{fig:Drude bounds}. 
	Negative Q-factor values are irrelevant for our study, and as such they must be ignored when they occur in our bounds. 

	\begin{figure*}[ht!]
		\centering
		\includegraphics[width=0.9\textwidth]{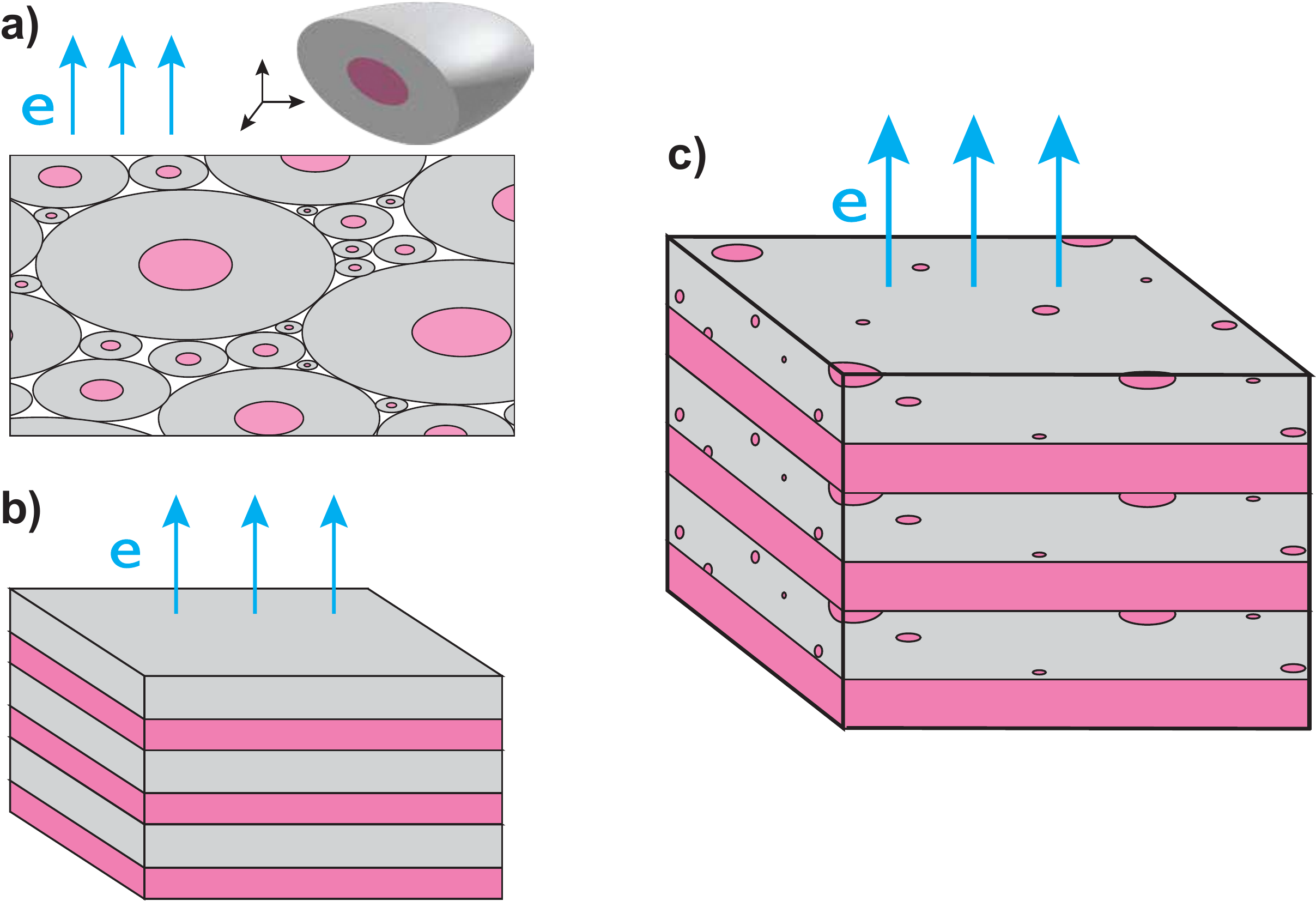}
		\caption{Optimal two-phase metamaterial microstructure designs: Pure phase with permittivity given by Drude model is shown in pink and the pure phase with constant permittivity is shown in gray. 
			a) Coated ellipsoid (CE1) assemblage corresponding to point A in Figure \ref{fig:Drude bounds}. 
			b) Laminate geometry (L), corresponding to point B in Figure \ref{fig:Drude bounds}.
			c) Doubly coated ellipsoid (DCE1) assemblage (Point C in Figure \ref{fig:Drude bounds}).
			In each of the designs, optimal $\Qnew$ is obtained in the direction of the applied electric field (seen in blue).
		}
		\label{fig:Optimal designs}
	\end{figure*}
	
	Further, we also find the optimal metamaterial microstructures that achieve these bounds.
	Specifically, we find that points on the upper horizontal bound are achieved by assemblages of confocal coated ellipsoids (CE1) with the core phase given by $\epsilon_1(\omega)$ \eqref{Drude model}.
	The effective permittivity of the coated ellipsoid (CE1) has two free parameters\cite{milton2002theory}; the depolarization factor and the volume fraction. 
	The depolarization factor is used to tune the resonance and the volume fraction can be varied within limits to trace points on the upper bound. 
	In Figure \ref{fig:Drude bounds} the yellow square markers denote the values from CE1. 
	Simple laminate geometries (L) attain only the extreme right point on the bounds (green triangle), when the layers of the laminate are normal to the direction of the electric field.
	Since, volume fraction is the only free parameter in the complex effective permittivity of laminates, it is varied to fix the pole and we get only one point.
	In the case when electric field is parallel to the layers, the effective permittivity of the laminate is an arithmetic mean of $\epsilon_1(\omega)$  and $\epsilon_2=1$, and consequently there is no resonance observed at any frequency.
	The lower bound in Figure \ref{fig:Drude bounds} is traced by assemblages of highly sensitive doubly coated ellipsoids (DCE1, shown by red circles) where the innermost and outermost phases are the same and given by $\epsilon_1(\omega)$. 
	The inner coated ellipsoid and outer coated ellipsoid are not restricted to have the same eccentricities, and hence  are more general than confocal doubly coated ellipsoids.
	Doing so provides us with 4 parameters; the depolarization factor of the inner coated ellipsoid, depolarization factor of the outer coated ellipsoid, volume fraction of the Drude phase, and volume fraction of the core phase. 
	These parameters can be varied so that the resulting effective permittivity function $\eff(\epsilon_1,1)$ matches any formula that defines the bounds (see, Section 18.5 in Milton\cite{milton2002theory}). 
	As such, the assemblages of doubly coated ellipsoids necessarily attain values on the bound. 

	\begin{figure*}[ht!]
		\centering
		\includegraphics[width=\textwidth]{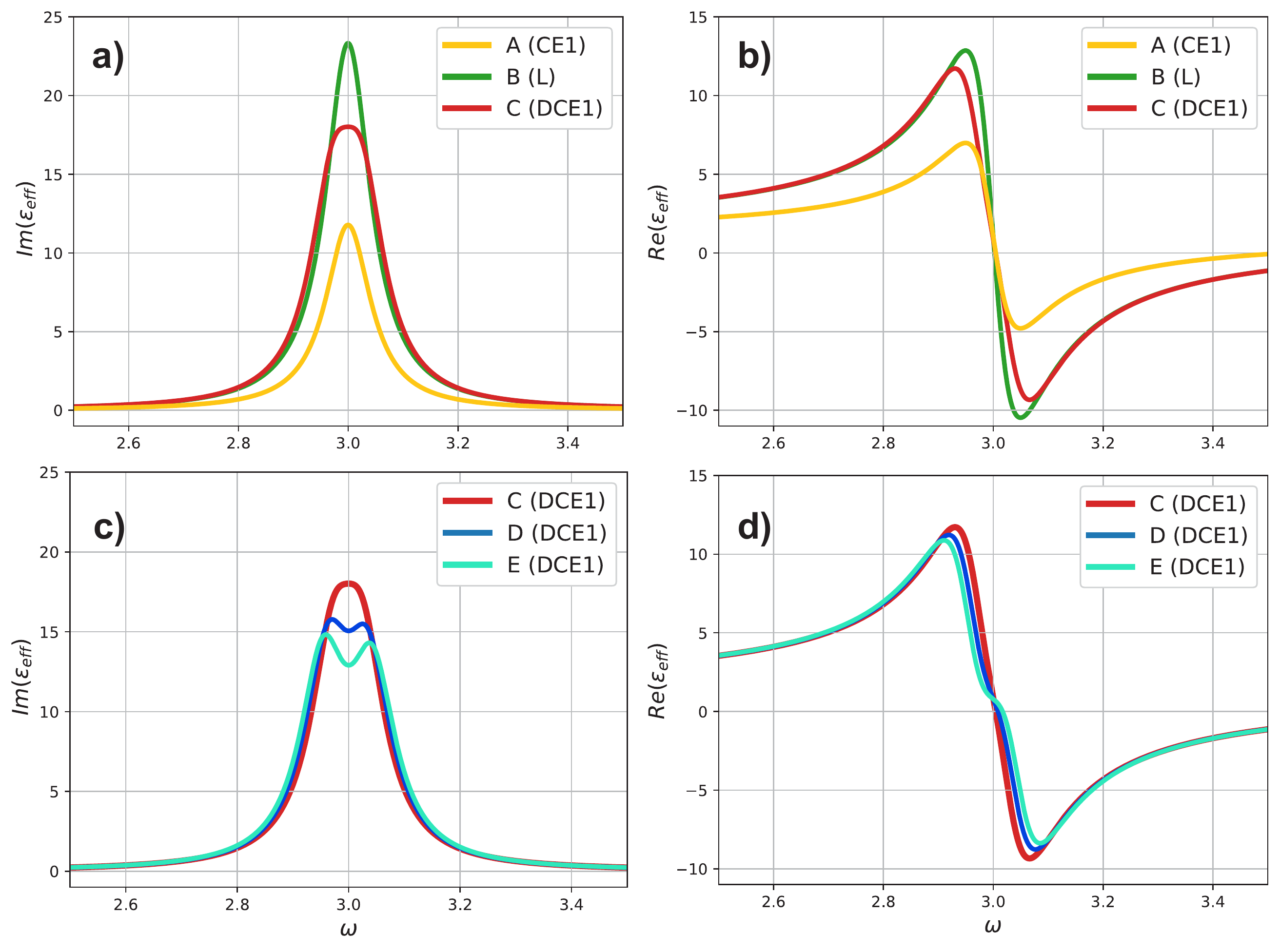}
		\caption{Plots of  $\eff$ vs. $\omega$, associated with the optimal metamaterial designs.
			a) $\Im(\eff)$ vs. $\omega$, and b) $\Re(\eff)$ vs. $\omega$, for the points A (coated ellipsoid, CE1, shown here by yellow line), B (laminate, L, shown by green line), and C (doubly coated ellipsoid, DCE1, shown by red line) from Figure \ref{fig:Drude bounds}.
			c), and d) show plots for $\Im(\eff)$ vs. $\omega$, and $\Re(\eff)$ vs. $\omega$, respectively, for three different doubly coated ellipsoid geometries that achieve the points C, D, and E on the bounds in Figure \ref{fig:Drude bounds}.
			Note, the behavior of $\Im(\eff)$ in c), where we can see a small local minimum (blue and cyan colored curves) at $\omega_R=3$.
		}
		\label{fig:respone}
	\end{figure*}
	
	In Figure \ref{fig:Optimal designs}, we present schematic drawings of the optimal metamaterial designs corresponding to three specific points A, B, and C on the bounds in Figure \ref{fig:Drude bounds}.
	In each of the subfigures, $\epsilon_1(\omega)$ is shown in pink color, and the constant phase $\epsilon_2 = 1$ is shown in gray color. 
	Point A corresponds to a coated ellipsoid assemblage depicted in Figure \ref{fig:Optimal designs}\textcolor{red}{a}.
	Point B corresponds to a material with laminate geometry which is shown in Figure \ref{fig:Optimal designs}\textcolor{red}{b}.
	Most interesting, is Point C that corresponds to a limiting case of doubly coated ellipsoid geometry as described before, with the following parameters: 
	depolarization factors of the inner and outer coated ellipsoid are  $0.5614$ and $1$, respectively, the volume fraction of the outermost phase $=0.4374$, and volume fraction of the core phase $=0.0002363$.
	Thus, the ellipsoidal inclusions only occupy an extremely small volume fraction, but  are significant due to resonance effects. 
	The schematic drawing seen in Figure \ref{fig:Optimal designs}\textcolor{red}{c}, depicts this laminate geometry with the constant phase (outermost layer of DCE1) forming one of the laminates, and the second laminate being formed from a very dilute assemblage of coated ellipsoids (inner coated ellipsoid of DCE1).
	For each of these designs, $\Qnew$ values attaining the bounds are obtained when the electric field is applied in the direction shown by the blue arrows.

	Next, we obtain the response of $\eff(\omega)$ with respect to $\omega$ for the specific geometries indicated by the points A (CE1), B (L),  C (DCE1), D (DCE1), and E (DCE1) in Figure \ref{fig:Drude bounds}.
	Figures \ref{fig:respone}\textcolor{red}{a}-\ref{fig:respone}\textcolor{red}{b}, show the plots for  $\Im(\eff)$ vs. $\omega$ and $\Re(\eff)$ vs. $\omega$, respectively, for three different geometries given by points A, B, and C.
	We observe that despite the large difference in the $\Im(\eff)$ values of  points A (CE1, seen in yellow) and B (L, seen in green), they have the same $\Qnew$. 
	Figures \ref{fig:respone}\textcolor{red}{c-d} show a similar comparison, but for three similar doubly coated ellipsoid geometries that attain points C, D, and E on the lower bound.
	We make an interesting observation here with regards to the values of $\Im(\eff)$ at the resonant frequency $\omega=\omega_R$. 
	As we move from point C towards point E, we observe the values of $\Im(\eff)$ at resonance frequency undergo a transition and exhibit a small local minimum  at the center frequency of the bandwidth (shown by blue and cyan colored curves in Figure \ref{fig:respone}\textcolor{red}{c}).

	\begin{figure*}[ht!]
		\centering
		\includegraphics[width=\textwidth]{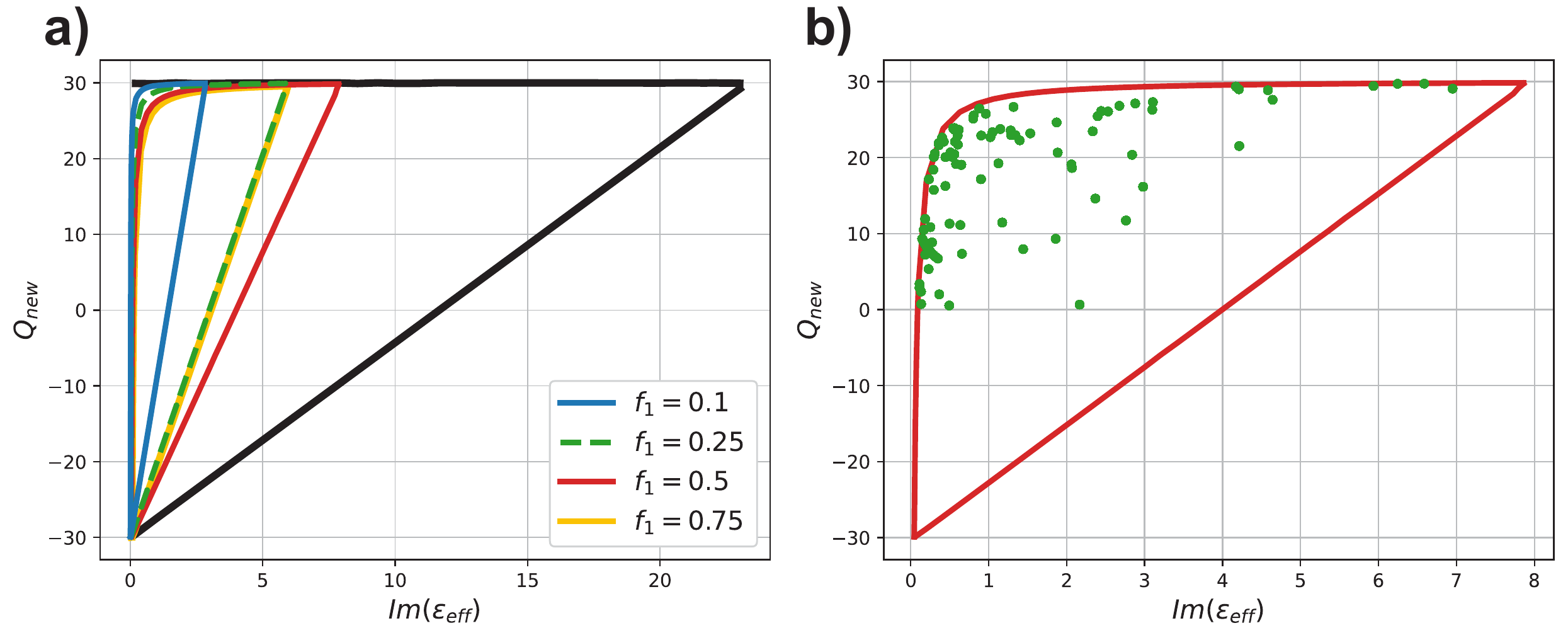}
		\caption{Bounds on $\Qnew$ for 3-d, isotropic, and fixed volume fraction 2-phase quasi-static metamaterial resonators at the resonance frequency $\omega_R=3$ with $\epsilon_1(\omega)$ given by the Drude model.
			a) $\Qnew$ bounds for volume fractions $f_1=0.1$ (blue line), $f_1=0.25$ (dashed green line), $f_1=0.5$ (red line), $f_1=0.75$ (yellow line) are superposed for comparison with the volume fraction independent bounds (solid black curve). 
			b) $\Qnew$ values obtained from Schulgasser laminates of doubly coated ellipsoids with the outer coated ellipsoid forming a prolate spheroid are plotted (green circles). 
			The solid red line in b) denotes the bounds for $f_1=0.5$. Some of the doubly coated ellipsoid geometries are found to be almost optimal
		}
		\label{fig:Drude Isotropic bounds}
	\end{figure*}

	Similarly, bounds on $\Qnew$ can be obtained for the case when the permittivity of the pure phase $\epsilon_1(\omega)$ is given by a Lorentz model.
	In the SI, we present plots for $\Qnew$ vs. $\Im(\eff)$ for 3 different frequencies; one, at the resonance frequency of the pure phase, two, near the resonance of the pure phase, and three at a frequency away from the resonance of the pure phase (see, Section \textcolor{red}{S4} in SI).
	These results show that the region enclosed by the $\Qnew$ bounds can be non-convex.

	We now consider the problem of obtaining bounds on $\Qnew$, when the 2-phase metamaterials are three-dimensional (3-d), isotropic, and have fixed volume fraction $f_1$ for the pure phase $\epsilon_1(\omega)$, which is again given by the Drude model, and the second phase is chosen to be $\epsilon_2 =1$.
	To obtain bounds that include the volume fraction and the fact that the microstructure is isotropic, we let $f_2=1-f_1$, be the volume fraction of phase 2 and introduce the function\cite{milton1985thermal},
	\begin{equation}
		\widetilde{\varepsilon}_\mathrm{eff}(\epsilon_1, \epsilon_2) = -\frac{1}{2}f_2\epsilon_1 - \frac{1}{2}f_1\epsilon_2 + \frac{f_1f_2(\epsilon_1-\epsilon_2)^2}{2[f_1\epsilon_1+f_2\epsilon_2-\eff]},
	\end{equation}
	(related to the so called Y transform\cite{milton1981bounds}. See, Chapters 19 and 20 in Milton\cite{milton2002theory}) that has basically the same analytic properties, and hence is subject to basically the same bounds, as $\eff(\epsilon_1,\epsilon_2)$.
	In the isotropic case bounds on $\eff$ with known volume fraction  were first obtained by Bergman and Milton\cite{milton1980bounds,PhysRevLett.45.148.2,milton1981bounds,milton1981bounds-b}, and were recently improved by Kern, Miller and Milton\cite{kern2020tight}. 
	We seek bounds that also involve $\deriv{\eff(\omega)}{\omega}$ (see SI  Section \textcolor{red}{S4} for more explanation).
	
	Bounds on $\Qnew$ for all possible values of $\Im(\eff)$ are obtained for four different volume fractions, $f_1 = 0.1, 0.25, 0.5,$ and $0.75$, for a resonance frequency of $\omega_R=3$. 
	Figure \ref{fig:Drude Isotropic bounds}\textcolor{red}{a} shows all the 3-d, isotropic, fixed volume fraction bounds superposed on top of the bounds shown in Figure \ref{fig:Drude bounds} for comparison, as they are all evaluated at the same frequency $\omega_R=3$. 
	The plots show that the area of the region occupied by the bounds does not monotonically increase as the volume fraction is increased, which is clear since the bounds for $f_1=0.5$ (solid red curve) occupies a larger region than the bounds for $f_1=0.75$ (solid yellow curve), and the bounds for $f_1=0.25$ (dashed green curve).

	Here too, we find some optimal isotropic, fixed volume fraction metamaterial designs that attain certain points on the bounds by using the Schulgasser\cite{schulgasser1977bounds} lamination technique to construct the isotropic effective permittivities.
	Given an anistropic effective tensor $\boldsymbol{\varepsilon}_\mathrm{eff}$, he showed that one could obtain an isotropic material with permittivity $\mathrm{Tr(\boldsymbol{\varepsilon}_\mathrm{eff})}/3$.
	
	The almost optimal geometries are Schulgasser laminates of assemblages of doubly coated ellipsoids with the outer coated ellipsoid forming a prolate spheroid, while there is no special form of the inner coated ellipsoid (see SI for parameter values). 
	Figure \ref{fig:Drude Isotropic bounds}\textcolor{red}{b}, shows the  $\Qnew$ values (green dots) obtained by doubly coated ellipsoids with outer coated ellipsoid taken to be prolate spheroids  for a volume fraction of $f_1=0.5$, with  some microstructures apparently attaining points on the bound.
	
	Similar, bounds on $\Qnew$ can be obtained for 3-d, isotropic, fixed volume fraction materials using Lorentz model for $\epsilon_1(\omega)$. The associated results are presented in the supplementary information. 
	Note that, when one of pure phases is Lorentzian, $\eff$ may have resonances due to both resonance of the pure Lorentzian phase, and resonances due to the composite microgeometry. 
	Instead, if the pure phase is given by the Drude model \eqref{Drude model}, then $\eff$ has resonances due to microstructure geometry only.

	In conclusion, our work provides limits on the quality of resonances that can be achieved in 2-phase quasi-static metamaterial resonators.
	Such resonances are important in nano-photonics and optics.
	Optimal metamaterial microstructures have been identified that achieve these limits. 
	It will be interesting to see then, how these theoretical and numerical results compare against any experiments done to validate the results in the quasi-static regime or invalidate them at higher frequencies.
	Since, this work provides bounds on $\Qnew$ for isotropic metamaterials too, it can be applied to design directional and omni-directional resonance responses in metamaterials.

	
	See supplementary information document for additional details related to this work.

	The authors are grateful to the National Science Foundation for support through grant DMS-2107926. 
	We thank Mats Gustafsson for suggesting a connection between Q-factor and the derivative $\Re\left(\deriv{\eff(\omega)}{\omega}\right)$ at resonance. 
	This formed the basis for our formula for $\Qnew$.
	

	
	\section*{Author Declarations}
	\subsection*{Conflict of Interest}
	The authors have no conflicts to disclose.
	
	\subsection*{Authors' Contribution}
	K. J. Deshmukh and G.W. Milton contributed equally on this paper.
	\section*{Data Availability}
	Aside from the data available within the article, any other data are available from the corresponding author upon request. 
	\newpage
	\section*{References}

	\bibliography{main}
\end{document}